\documentstyle[preprint,aps]{revtex}

\title{\bf Charm Production in Flux Tubes}

\author{C.E.~Aguiar,$^{1,2}$ T.~Kodama,$^1$
        R.A.M.S.~Nazareth,$^1$ and G.~Pech$^3$}

\address{$^1$Instituto de F\'\i sica,
         Universidade Federal do Rio de Janeiro, \\
         C.P. 68528, 21945-970, Rio de Janeiro, Brazil.}

\address{$^2$European Centre for Theoretical Studies in
         Nuclear Physics and Related Areas, \\
         Villa Tambosi, I-38050, Trento, Italy.}

\address{$^3$Instituto de F\'\i sica,
         Universidade do Estado do Rio de Janeiro, \\
         20550-013, Rio de Janeiro, Brazil.}

\date{July 17, 1995}

\preprint{ECT*/Jul/95-03}

\draft

\begin{document}

\maketitle

\begin{abstract}
We argue that the non-perturbative Schwinger mechanism may play an
important role in the hadronic production of charm.  We present a flux
tube model which assumes that the colliding hadrons become color
charged because of gluon exchange, and that a single non-elementary
flux tube is built up as they recede. The strong chromoelectric field
inside this tube creates quark pairs (including charmed ones) and the
ensuing color screening breaks the tube into excited hadronic
clusters.  On their turn these clusters, or `fireballs', decay
statistically into the final hadrons. The model is able to account for
the soft production of charmed, strange and lighter hadrons within a
unified framework.
\end{abstract}


\newpage


\section{Introduction}

Charm production in hadronic and nuclear collisions is presently
a subject of considerable interest, as charmed particles are
expected to be copiously created in relativistic heavy ion
reactions at BNL's Relativistic Heavy Ion Collider (RHIC)
and CERN's Large Hadron Collider (LHC).
On one side, high energy electrons and muons coming from
semileptonic decays of these charmed particles will provide a
significant `background' which may obscure dilepton signatures
of a quark-gluon plasma \cite{Vogt}.
On the other side, charm production itself may be an interesting
probe of the QCD plasma due to the mass scale of $c$~quarks.
It has been suggested that the observation of enhanced charm
production in heavy ion collisions could provide a way for
measuring the temperature of a hot gluon plasma \cite{Shuryak}
or the thermalization time of the initial partonic system
\cite{Muller}.

In order to study charm production in relativistic heavy ion
reactions one needs to have a good understanding of the
production mechanisms operating in nucleon-nucleon collisions.
Only in this case one can make meaningful extrapolations from
hadronic to nuclear collisions, and use them to identify
new phenomena associated with dense matter formation.
Charm production in hadronic collisions has been
extensively investigated with perturbative QCD.
Of particular interest for such studies are the recent
calculations of heavy quark hadroproduction carried out to
order $\alpha_s^3$ by various authors \cite{heavyq}.
These next-to-leading-order calculations seem to describe
very well several experimental features of bottom production,
but when applied to charm they give a $c \bar c$~cross section
which depends strongly upon the choice of the
renormalization-factorization scale.
This renders it difficult to assess from direct comparisons
with data whether hard processes really represent the dominant
source of charm in hadronic collisions.
Furthermore, there are important features of the
experimental data which are not reproduced by these
perturbative calculations.
Some examples are the small back-to-back $p_T$ correlations
found in the charm-pair distribution \cite{correl}, or the
enhancement of leading charm production observed at large
$x_F$ \cite{leading}.

This leaves open the possibility that non-perturbative
mechanisms have an important role in charm production.
In fact, the relatively low mass of the $c$-quark places
the hadronic production of charm on the border between
perturbative and non-perturbative phenomena.
Therefore, approaches from the non-perturbative side may
prove instructive.
Such approaches are usually in the form of phenomenological
models.
Although less rigorous,  they provide useful physical images
whose parameters, like the string constant or the vacuum
pressure, should be object of more fundamental theories.
However, the usual mechanisms for soft particle production do
not give a good description of charm hadroproduction.
The commonly used string models predict negligible charm
yields, because the tension of a quark-antiquark flux tube is
too small for it to break into a $c \bar c$~pair \cite{lund}.
Statistical (hydrodynamical) models also do not predict any
significant charm production, as typical fireball temperatures
are much smaller than the charm mass.

It is then somewhat surprising to find out that by properly
blending the string and statistical approaches one is able
describe very well the hadroproduction of charmed particles.
Such a hybrid approach to particle production (we call it
the ``firetube'' model) was developed in Ref.\cite{ftube} to
study the rapidity and transverse momentum distributions of
pions and nucleons in hadronic collisions.
In the present paper we show how the firetube model can be
extended to describe charm and strangeness production.

The general idea of the firetube model is as follows.
We assume that the colliding hadrons exchange soft gluons
when they pass through each other, and as a result acquire
color charges.
The receding hadrons then become connected by a flux tube
which confines the strong chromoelectric field created by
these charges.
Quark pairs (including $c \bar c$) are created by the
chromoelectric field via the Schwinger mechanism, and the
resulting color screening breaks the flux tube into lumps of
highly excited hadronic matter (``fireballs'') which
subsequently decay thermally into the observed hadrons.
The firetube model shares many aspects with the usual models
based on the fragmentation of a classical string, such as the
Lund model \cite{lund}.
However, there exist basic differences.
First, because of the gluon exchange, in the firetube model
the string tension can be much larger than that of an
elementary string between a quark-antiquark pair.
Second, the final hadrons come from the thermal decay of
fireballs.
These points bring some new aspects into the mechanism of
hadron production.
For example, in a common string fragmentation model, the
hadrons are produced directly from the break up of a
quark-antiquark string via the Schwinger mechanism.
Therefore, in such a framework the so-called $K/\pi$
ratio is closely related to the string constant $\kappa$.
In contrast to this, the mechanism for pion production
in the firetube model is essentially the thermal decay
of fireballs, and there is no contradiction here between the
small $K/\pi$ ratio and a large hadron-hadron string tension.
Furthermore, if the string constant is as small as the standard
value $\kappa \simeq 1$~GeV/fm of a $q \bar q$-string, the
Schwinger mechanism does not produce any significant
amount of charm, as we have already mentioned.
However, the Schwinger pair-creation rate is very sensitive to
the value of $\kappa$, and the larger string constants found in
the firetube model drastically change the picture, providing
a mechanism for abundant non-perturbative production of charm.

The aim of this paper is to investigate such mechanism.
A brief description of the firetube model is given in
Sec.~\ref{f-m}, and results for pion production are presented.
In Sec.~\ref{c} we discuss charm production in the framework of
this model.
The total charm cross section is calculated for proton-proton
collisions and the result compared to experimental data.
The longitudinal and transverse momentum distributions of
charmed mesons and baryons are also obtained and compared to
measurements.
In Sec.~\ref{s} we investigate strangeness production and
discuss $K$ and $\Lambda$ spectra.
Finally, Sec.~\ref{fin} is used for some further comments and
conclusions.


\section{The Firetube Model}
\label{f-m}

When two hadrons collide, several sea partons (assumed to be gluons
here) may be exchanged between them.
As a result, these hadrons become colored objects linked by a
flux tube.
Let $S$ be the cross section of this tube, which should be of the
order of the geometrical size of the colliding hadrons. Let also $Q$
be the color charge at the end points of the flux tube, measured
in units of the elementary color charge $q_o$ of a quark
($q_o=\sqrt{4/3} \, g_s$, where $g_s$ is the QCD coupling constant).
The chromoelectric field $E$ inside the tube can be calculated
from Gauss' law to be $ E = q_o Q / S $.
Placed in such a field, a quark whose color charge points into the
$E$ direction in color space (we omit $SU_3$ algebra indices for
simplicity) will experience a force $q_oE=q_o^2Q/S$.
Also, the string constant $\kappa$ is related to the field energy
density by $\kappa =E^2S/2$. From this we get
\begin{equation}
\kappa = \kappa_o Q^2
\end{equation}
\begin{equation}
q_oE = 2 \kappa_o Q
\end{equation}
where we have defined $\kappa_o = q_o^2/(2S)$.
We identify $\kappa_o$ as the string constant of an ``elementary''
($Q=1$) color triplet string.
It is worthwhile to mention that the scaling relations we
are using --- $S \sim Q^0$, $E \sim Q^1$, $\kappa \sim Q^2$ ---
are  very different from what one obtains with the MIT bag
model. In this case the balance between the field energy
density and the vacuum pressure leads to an increase of the
flux tube cross section with the charge, $S \sim Q^1$, and
the chromoelectric field and string constant scale as
$E \sim Q^0$, $\kappa \sim Q^1$ .
However, lattice QCD calculations of flux tubes generated
by sources in different representations of the color group
do predict that the scaling of $S$, $E$ and $\kappa$ with $Q$
is the one we have used above \cite{lattice}.

The end-point charge $Q$ is not necessarily the same for every
collision. It fluctuates because different numbers of gluons can be
exchanged, and also because of the $SU_3$ addition of color charges.
Assuming that each gluon exchange is a step of a random walk in
color space \cite{Biro}, it can be shown that the charges generated
on the hadrons after the exchange of $n$ gluons distribute sharply
around the mean value $(3/2)\sqrt{n}$, where the 3/2 factor accounts
for the gluonic octet charge.
Thus, $\kappa$ and $q_oE$ fluctuate around
\begin{equation}
\kappa \simeq {9\over 4}\kappa _o n \;,\qquad
q_oE\simeq 3\kappa _o\sqrt{n} \;,
\end{equation}
showing that the exchange of even a modest number of gluons gives
rise to string constants and chromoelectric fields significantly
larger than the elementary ones.

In order to determine completely the statistical distribution of the
color charge $Q$ we must know the probability of having $n$
gluons exchanged in a collision.  We assume that this is given
by a truncated Poisson distribution,
\begin{equation}
P_n \sim \nu ^n/n! \;, \qquad n\geq 1 \;,
\end{equation}
where $\nu$ is a parameter related to the average number of gluons by
$\overline{n} = \nu/(1-e^{-\nu})$.

The constant chromoelectric field $E$ produces quark-antiquark
pairs inside the flux tube by a process similar to the Schwinger
mechanism of electron-positron creation in QED.  The $q \bar q$
production rate per unit volume per unit time can be calculated (in
the Abelian approximation) from Schwinger's formula
\begin{equation}
\label{Schwinger}
{\cal R}_q =  \frac{(q_oE)^2}{4\pi^3}
\sum_{n=1}^\infty \frac{1}{n^2}\exp
\left\{ -\frac{n\pi {m_q}^2}{q_oE}\right\}
\end{equation}
where $m_q$ is the quark mass.
Corrections for the final state interaction of the
$q \bar q$ pair can be introduced into this formula by
taking $q_oE=2\kappa_o Q - \kappa_o$.
In our calculations we also corrected Eq.(\ref{Schwinger}) for
the finite transverse size of the flux tube, using the
semiclassical formula of Ref.\cite{Vautherin} (see their eq.27).
For the cases of interest to us, the simple semiclassical
result is not very different from the  more precise
correction of Ref.\cite{Brink}.
In principle we should also correct the Schwinger formula
for the finite length of the flux tube \cite{Wong}.
But it has been shown \cite{Vautherin} that if the tube end-points
are moving rapidly (which is our case) such corrections are
suppressed by relativistic dilation effects. For this reason we
have neglected longitudinal size corrections.

The quarks created by the chromoelectric field have their color
charges aligned with $E$ in $SU_3$ space (the other orientations will
be ignored as they have a much smaller production rate
\cite{Biro,Casher})
and tend to screen the end-point charges $Q$, providing a mechanism
for the flux tube fragmentation.  Although other processes such as
collective instabilities of the vacuum can be invoked to explain the
break up of the tube into fireballs, we will assume that pair creation
gives the dominant mechanism.  This has the advantage of allowing for
a simple estimate of the firetube fragmentation rate per unit length
per unit time as
\begin{equation}
\omega \approx {S\over Q} \sum_q {\cal R}_q
\end{equation}
where the sum extends over the quark flavors ($u$, $d$ and $s$
in practice). This is, admittedly, a rough treatment of screening
effects, but we did check that the typical break-up time we obtain
in this way is comparable to the collapse time of the color field
calculated with quark transport models \cite{Wilets}.

The firetube model has a relatively small number of parameters.
For proton-proton collisions we take
$\overline{n}=2.0$, $\kappa_o=1$~GeV/fm and $S=1.5\hbox{ fm}^2$.
Note that these values are quite reasonable: $\kappa_o$ is the usual
$q \bar q$ string constant \cite{lund,Casher}, and the firetube
radius $R = \sqrt{S/\pi} \simeq 0.7$~fm is almost the same as
that of the proton.  A few other parameters define the minimum
fireball mass (1 GeV), regulate the behavior of
the leading particles, and determine how the effective temperature
and longitudinal expansion rate of a fireball depend upon its mass
\cite{ftube}.  For the quark masses we use the constituent values
$m_u=m_d=300$~MeV, $m_s=450$~MeV and $m_c=1.5$~GeV \cite{Casher}.

Having defined all this, we can perform Monte Carlo simulations of
the firetube formation, its fragmentation into fireballs, and the
thermal decay of these into the observed hadrons \cite{ftube}.
A typical result of such calculations is presented in Fig.1, where
we show the rapidity distribution of charged particles (mostly pions)
produced in proton-proton collisions at $\sqrt{s}=20$~GeV and 53~GeV.
We see that the model calculation is in good agreement with the
experimental data \cite{charged} at both energies.
Results of similar quality are obtained for the transverse momentum
distributions.


\section{Charm Production in the Firetube Model}
\label{c}

As it can be seen from Schwinger's formula, Eq.(\ref{Schwinger}), the
production rate of charmed quarks is extremely sensitive to the value
of $q_oE$.  For example, if $q_oE$ were of the order of 1~GeV/fm as
usually quoted in the string model, the dominant exponential factor
in Eq.(\ref{Schwinger}) would be of the order of 10$^{-15}$, and no
reasonable interaction volume and time scale for hadron-hadron
collision would account for the observed charmed particle production
cross section.  On the other hand, in our model the average value of
$q_oE$ is approximately 3~GeV/fm and the average value of the
dominant exponential factor for charm becomes as large as 10$^{-4}$,
for the parameter values given above.

In order to obtain the charm production cross section we proceed as
follows.  First, we calculate the available space-time volume $VT$ as
the total area swept by the flux tube in the space-time plane times
the tube cross section $S$. This is completely determined if the
fragmentation scheme of a firetube is specified. The total number
of charmed quarks is then twice the value $VT \times {\cal R}_c$.
In Fig.2 we show the total charm production cross section calculated
in this way, and compare it to experimental data
\cite{NA27,E743,NA25,E653}. The agreement is seen to be good,
specially if we have in mind that the same parameter set is used at
all energies. In our model, the energy dependence of the charm
production rate comes essentially from that of the space-time volume
$VT$ swept by the firetube, which increases asymptotically
as $\ln (s)$.

To calculate the momentum distribution of charmed particles,
we assume that the $c$~quarks produced by the mean color field
are distributed among the tube fragments with
probabilities which are proportional to the mass of each fireball.
The charmed particles are then emitted from the fireballs
following their longitudinal expansion and thermal decay,
in the same way as the other mesons \cite{ftube}.
In Fig.3 we show our calculation
for the $x_F$ distribution of $D$ mesons produced in proton-proton
collisions at $\sqrt{s}=27$~GeV (solid line) together with
experimental points \cite{NA27}.  The agreement is very satisfactory.
The transverse momentum distribution of charmed
mesons predicted by the model is also in a good agreement with
experimental data \cite{NA27}, as shown in Fig.4.

In addition to the particles coming from fireballs, hadronic
spectra also get a contribution from the leading particles.  The
mechanism for tube breaking into fireballs assumes a
fragmentation rate $\omega$ homogeneous in space-time.
There is no apriori reason for this to remain valid at the two
end-points which contain the valence quarks, because of the
different boundary conditions.
In fact, to reproduce the observed leading nucleon spectra we have
to require that the two proton-like extremities detach from the
firetube with a probability rate that is constant on their
world-lines.
When these hadrons separate from the firetube, there
exists a chance that they turn into charmed particles,
as the detaching mechanism should also be related to the
production of $q\bar q$ pairs.
Thus, the probability of having a charmed leading particle
can be estimated as ${\cal R}_c / \sum_q {\cal R}_q$.
We have calculated in this way the $x_F$ spectrum of leading charmed
baryons ($\Lambda_c$'s) produced in proton-proton collisions, and
the result is indicated by the dotted line in Fig.3.
Note that these calculations do predict a substantial leading
particle effect, as the $\Lambda_c$ spectrum is much harder
than the $D/\bar D$ distribution.
For pion-nucleon reactions, where detailed observations
of the leading charm effect have been performed \cite{leading},
our calculations reproduce very well the experimental results
\cite{next}.
We should mention that other non-perturbative effects, like
intrinsic charm \cite{intrinsic} or valons \cite{valon}, have
also been invoked to explain the leading particle behavior.

Another important feature of the firetube model is that we expect
no strong transverse momentum correlations in the $D-\bar D$ pair
spectra, as the charmed particles are emitted from fireballs which
decay statistically.
This seems to be consistent with most of the available data,
which show almost no back-to-back angular correlations in the
transverse plane \cite{correl}.

A final point concerns charm production in hadron-nucleus collisions.
Some preliminary extensions of our model indicate that the forward
($x_F>0$) charm production cross section is essentially proportional
to the target mass number $A$. However, for negative $x_F$ the charm
production rate seems to increases more rapidly than linear in $A$.


\section{Strangeness Production}
\label{s}

We have already seen that the large string tension arising in
hadronic collisions can accommodate both the pion and charmed
meson spectra quite satisfactorily.
At this point one may worry about strangeness production.
For simplicity we will assume here that strange particles are
created solely through the Schwinger mechanism, in the same way as
charmed particles.
This means we are ignoring thermal production of strangeness,
which in principle could take place inside the fireballs (for charm
this process is certainly negligible because of the large mass of
the $c$-quark).
In Fig.5 we show the $x_F$ spectrum of neutral kaons produced in
proton-proton collisions, calculated exactly as we did for the
$D$ mesons.
We see that the agreement of our results (shown as a solid line)
with the experimental data \cite{k-data} is very reasonable.
Transverse momentum distributions predicted by the model are also
in good agreement with experimental results.
The production of strange leading particles can also be calculated
along the same lines followed for the charmed ones.
Our result for the spectrum of leading $\Lambda$ baryons produced
in p-p reactions is shown in Fig.5 (dashed line).
The accordance with experimental data \cite{k-data} is again
quite good.


\section{Final Remarks}
\label{fin}

In this paper we have discussed the possibility that a
non-elementary flux tube is created between two colliding hadrons
because of gluon exchange.
This picture is consistent with charmed particles being produced
at large rates by the color field inside the tube, via the
Schwinger mechanism.
We developed a simple model based on this idea and showed that
it could account quantitatively for the production of charm in
proton-proton reactions.
The same model also describes very well the creation of lighter
hadrons such as pions and kaons.
It is a noteworthy aspect of our results that a successful
description of the $p_T$ and $x_F$ spectra of such different
particles as $\pi$'s, $K$'s and $D$'s can be achieved in a unified
manner.

An important issue refers to the competition between the mechanism
for non-perturbative charm production proposed here and perturbative
partonic processes.
Hard processes certainly contribute to the charm yield but, as
mentioned in the introduction, the present uncertainties in
perturbative calculations make it difficult to determine the size
of such a contribution.
Even though we have shown that non-perturbative production by the
color mean field can easily account for the observed charm cross
section, we do not claim here that this mechanism represents
the only significant source of charmed particles.
It should be clear that the precise values of our charm cross
section also suffer from uncertainties coming from parameter values.
For example, by adopting a slightly larger value for the mass of the
$c$-quark we would reduce our charm production cross section without
changing any of the model predictions for non-charmed particles.
The basic point we want to stress here is that, within a reasonable
set of parameters, our model shows that an important part of
the charmed particles produced in a hadronic collision may have a
non-perturbative origin.
So, from our point of view, soft and hard mechanisms may well be
both necessary in order to account for the bulk of charm production.
What is certain is that several experimental observations
\cite{correl,leading} cannot be described on purely perturbative
grounds, and require a significant contribution of soft mechanisms
in order to be understood.
In this paper we presented a consistent non-perturbative framework
that can describe not only these but most features of charm
production.
For the reasons discussed in the introduction, it is of interest
to investigate how the mean field mechanism for charm production
discussed here scales to heavy ion collisions.
Work along these lines is in progress.

\acknowledgments
This work has been supported partially by CNPq, FINEP
and FAPESP. G.~Pech and C.E.~Aguiar acknowledge CNPq Fellowships at
CBPF and ECT$^*$.



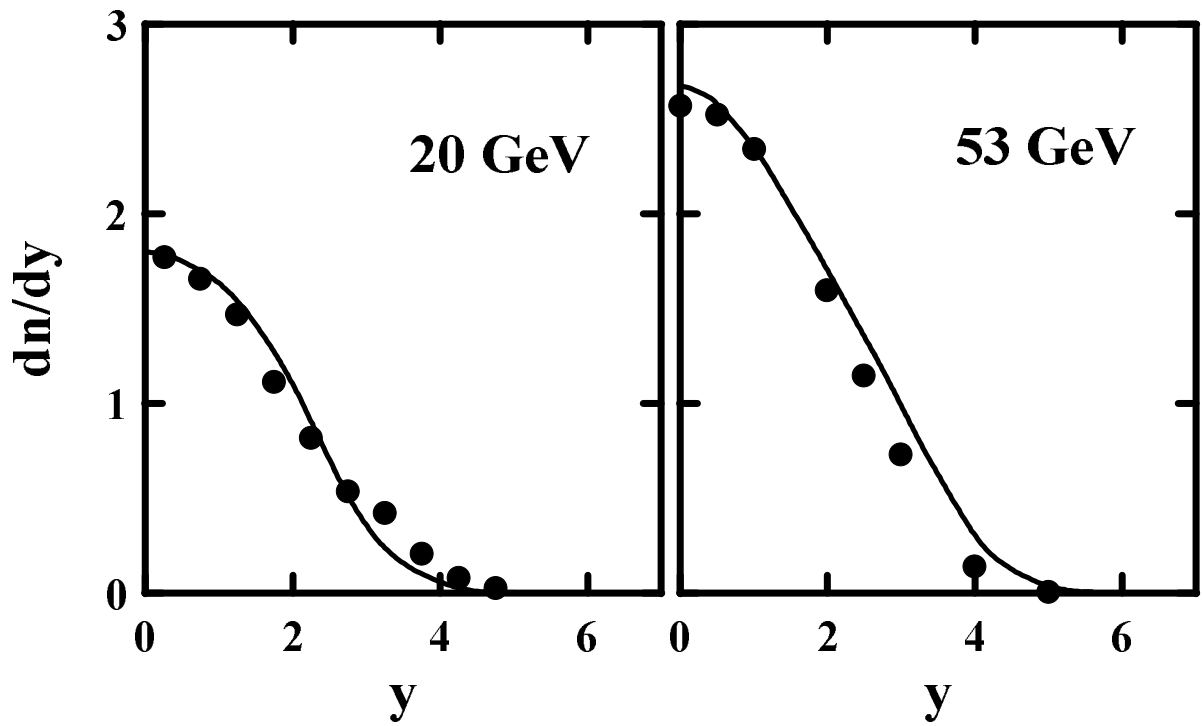
\begin{figure}
\caption[1]{
Rapidity distribution of charged particles produced in p-p
collisions at $\sqrt s$ = 20~GeV and 53~GeV.
Experimental data are from Ref.\cite{charged}.
The curves represent the result of Monte-Carlo calculations with
the firetube model.}
\end{figure}

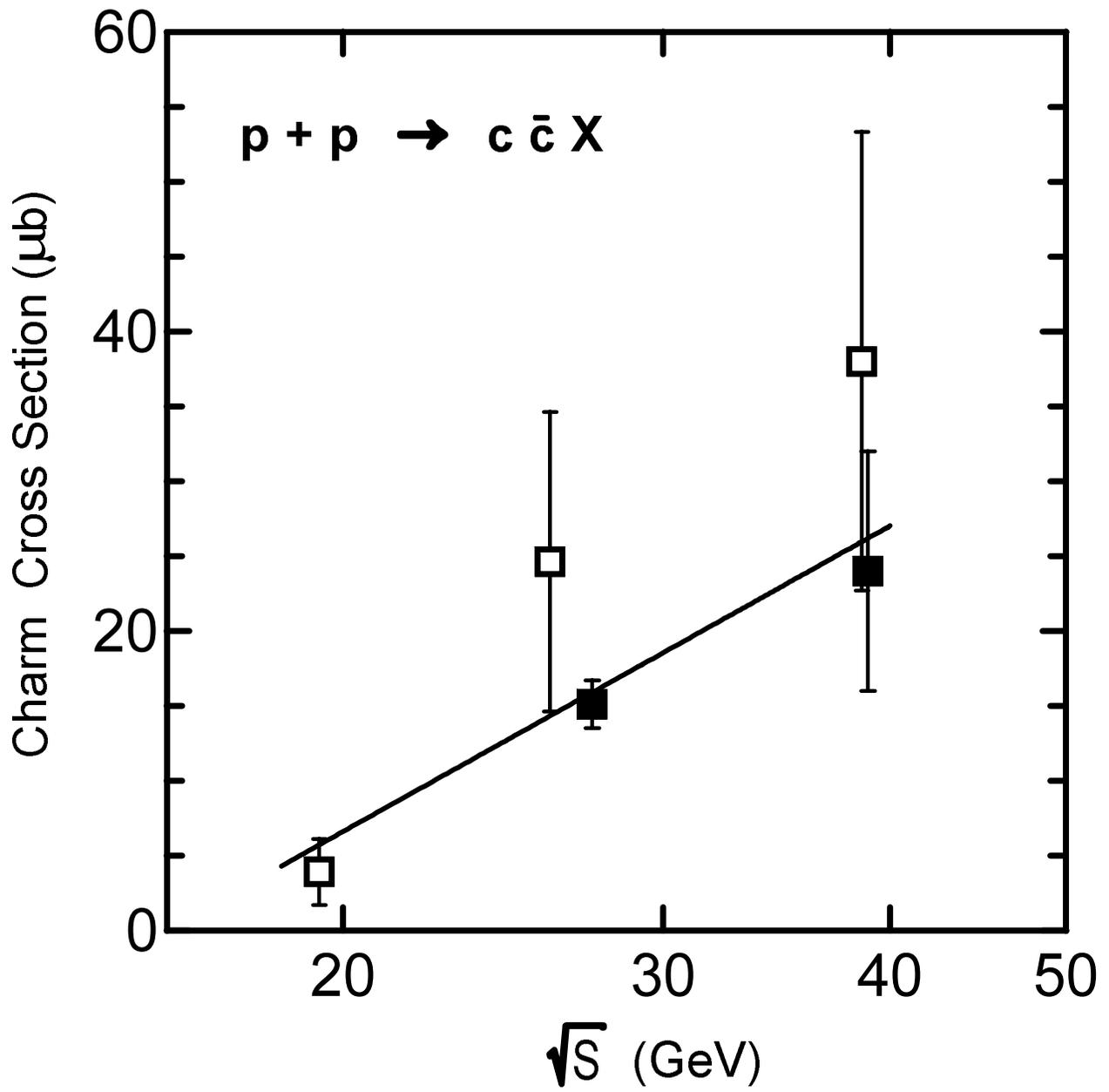
\begin{figure}
\caption[2]{
Total cross section for charm production in p-p collisions as a
function of c.m.~energy.
The line is the result of the firetube model.
Experimental points (solid squares) are from Refs.\cite{NA27,E743}.
The open squares refer to data obtained with proton-nuclei
reactions \cite{NA25,E653}.}
\end{figure}

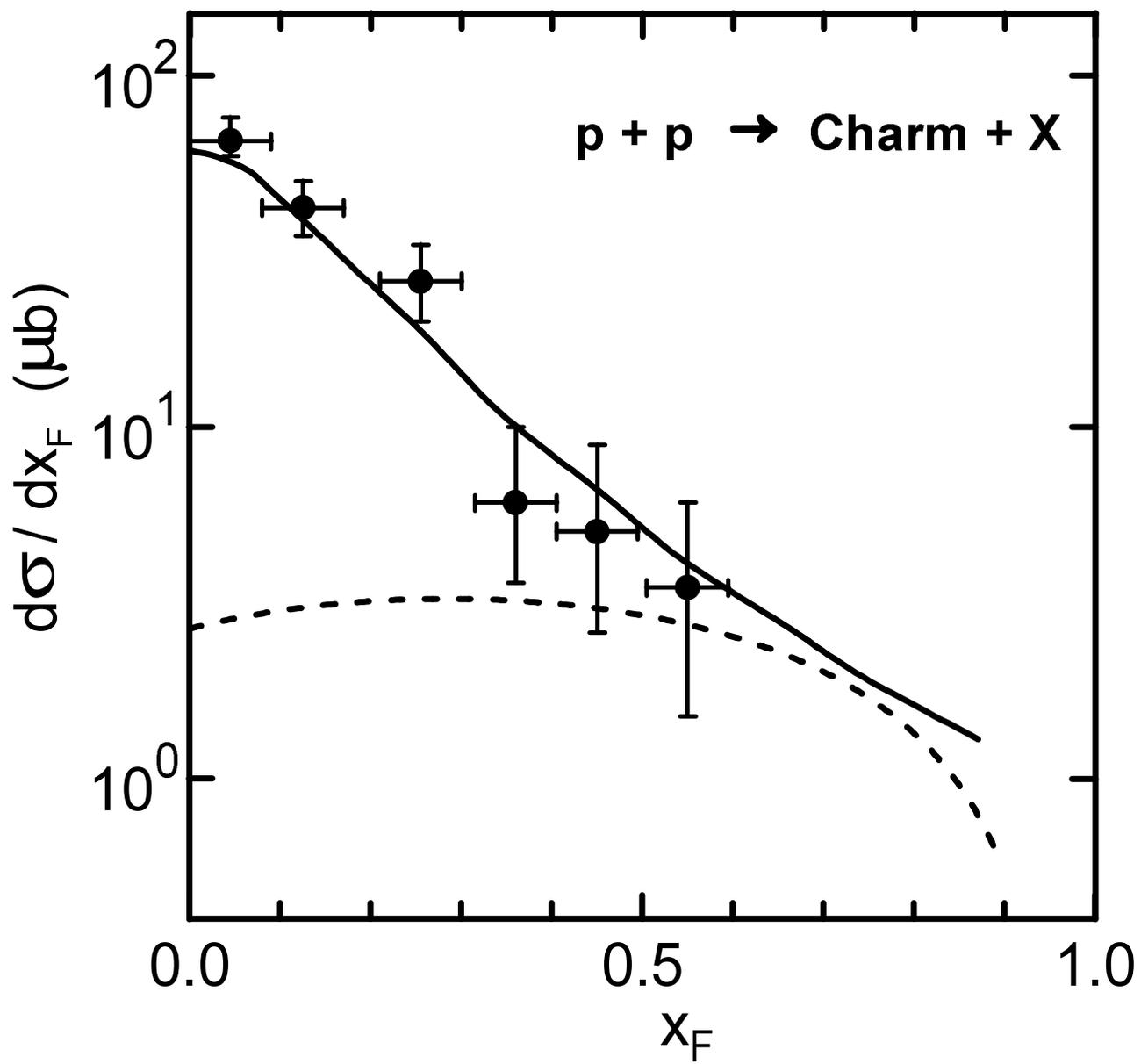
\begin{figure}
\caption[3]{
Feynman $x_F$ distribution of charmed particles created in p-p
reactions at $P_{inc}$ = 400~GeV/c.
The solid and dashed curves correspond to our calculations for
$D/\bar D$ mesons and $\Lambda_c$ baryons, respectively.
Data points for $D/\bar D$ are from Ref.\cite{NA27}.}
\end{figure}

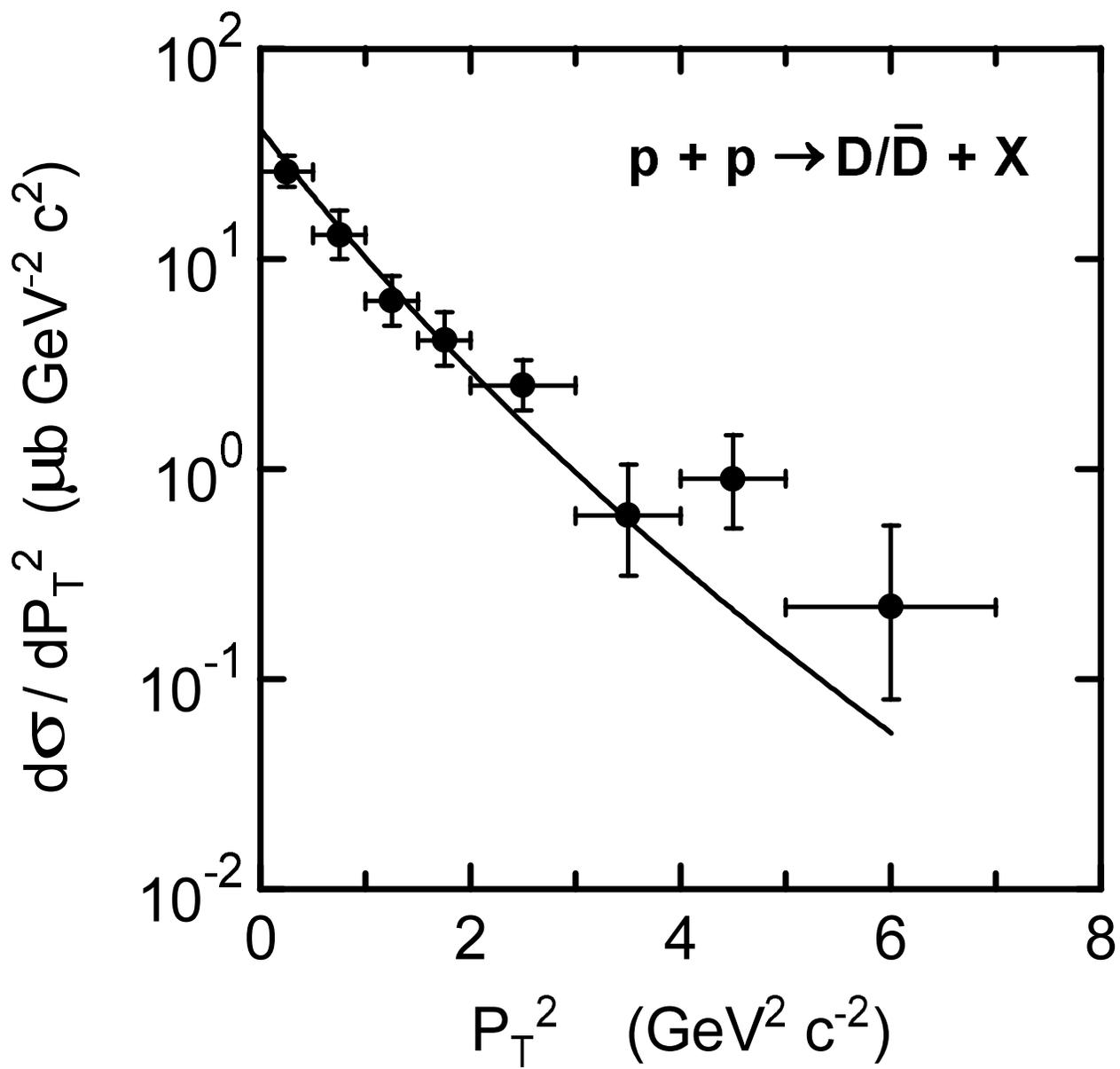
\begin{figure}
\caption[4]{
Transverse momentum distribution of charmed mesons for the
same reaction as in Fig.3.}
\end{figure}

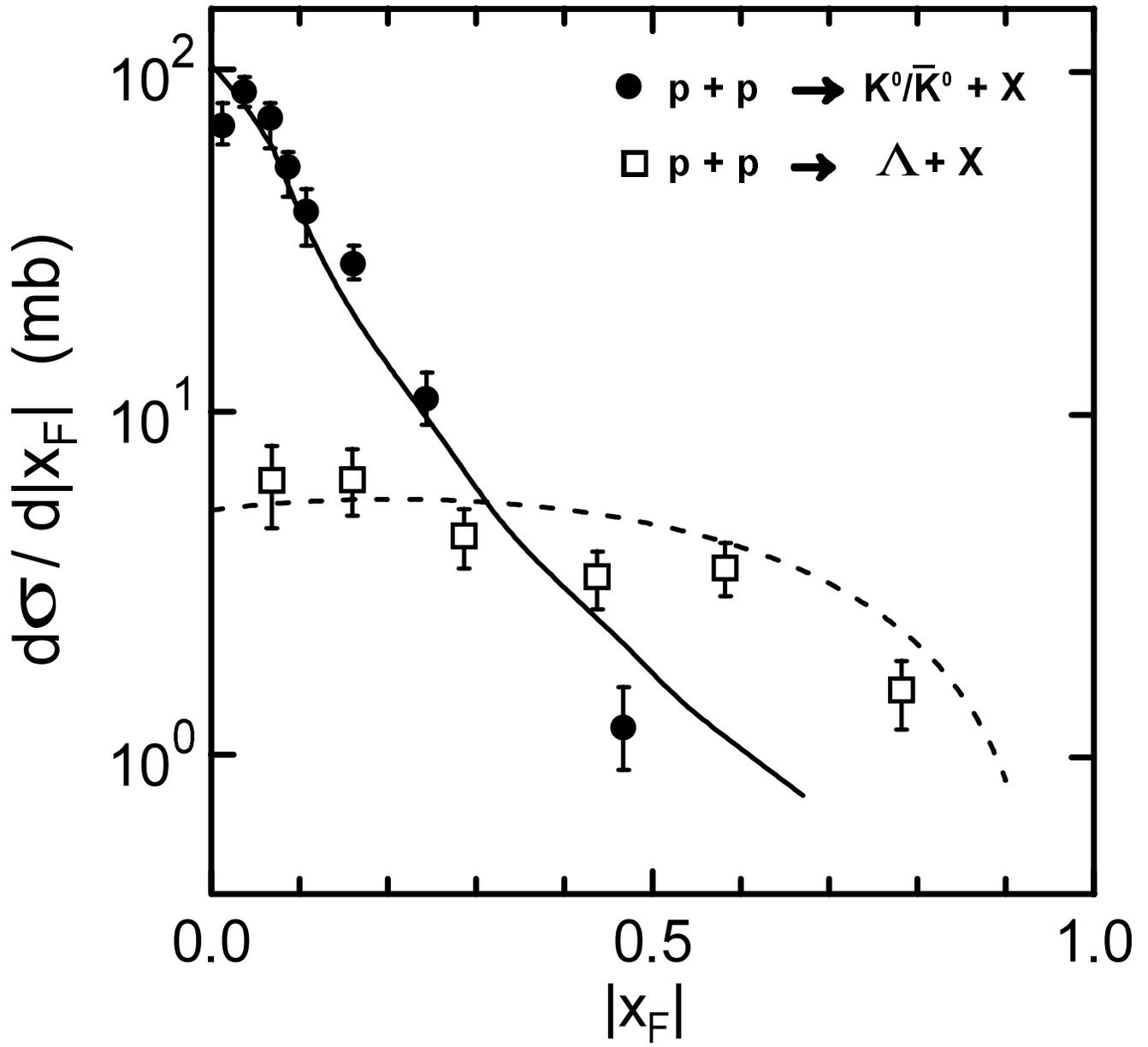
\begin{figure}
\caption[5]{
Feynman $x_F$ distribution of strange particles produced in p-p
collisions at $P_{inc}$ = 205~GeV/c.
The solid and dashed curves correspond to our calculations for the
$K^0/\bar K^0$ mesons and $\Lambda$ baryons, respectively.
Experimental data are from Ref.\cite{k-data}.}
\end{figure}

\end{document}